\documentclass[aps,prl,twocolumn,groupedaddress]{revtex4-1}  
\usepackage{dcolumn}   
\usepackage{bm}        
\usepackage{amsfonts}
\usepackage{wasysym}
\usepackage{amssymb}
\usepackage{graphicx}
\usepackage{caption}
\usepackage{subcaption}
\usepackage{booktabs}
\usepackage{array}
\usepackage{paralist}
\usepackage{verbatim}
\usepackage{subfig}
\usepackage{braket}
\usepackage{color}

\newcommand{\axion}{\ket{a}}

\newcommand{\beq}{\begin{equation}}
\newcommand{\be}{\begin{equation}}
\newcommand{\ee}{\end{equation}}
\newcommand{\bea}{\begin{eqnarray}}
\newcommand{\eea}{\end{eqnarray}}
\newcommand{\eeq}{\end{equation}}
\newcommand{\bef}{\begin{figure}}
\newcommand{\eef}{\end{figure}}

\newcommand{\kpc}{\text{ kpc}}
\newcommand{\half}{\frac{1}{2}}
\newcommand{\nn}{\nonumber}
\newcommand{\ti}{\times}
\newcommand{\mc}{\mathcal}

\def\z{\Theta}
\def\d{\Delta}

\def\Pga{P_{\gamma \to a}}

\begin{document}
\title{Galaxy Cluster Thermal X-Ray Spectra Constrain Axion-Like Particles}
\author{Joseph P. Conlon}
\email{joseph.conlon@physics.ox.ac.uk}
\affiliation{Rudolf Peierls Centre for Theoretical Physics, University of Oxford,\\Keble Road, Oxford, OX1 3NP, United Kingdom}
\author{M. C. David Marsh}
\email{M.C.D.Marsh@damtp.cam.ac.uk}
\affiliation{Department of Applied Mathematics and Theoretical Physics,\\
University of Cambridge, Cambridge, CB3 0WA, United Kingdom}
\author{Andrew J. Powell}
\email{Andrew.Powell2@physics.ox.ac.uk}
\affiliation{Rudolf Peierls Centre for Theoretical Physics, University of Oxford,\\Keble Road, Oxford, OX1 3NP, United Kingdom}

\begin{abstract}
Axion-like particles (ALPs) and photons inter-convert in the presence of a magnetic field.
At keV energies in the environment of galaxy clusters, the conversion probability can become unsuppressed for light ALPs.
Conversion of thermal X-ray photons into ALPs can introduce a step-like feature into the cluster thermal bremsstrahlung spectrum, and
 we argue that existing X-ray data on galaxy clusters should be sufficient to extend bounds on ALPs in the low-mass region $m_a \lesssim 1 \times 10^{-12}\,{\rm eV}$ down to $M \sim 7\times 10^{11}\, {\rm GeV}$, and that for $10^{11}\, {\rm GeV} < M \lesssim  10^{12}$ GeV light ALPs give rise to interesting and unique
 observational signatures that may be probed by existing and upcoming X-ray (and
 potentially X-ray polarisation) observations of galaxy clusters.
\end{abstract}


\maketitle
Galaxy clusters are the largest gravitationally bound objects in the universe and provide a powerful testing ground for theories of new physics. Recently, it has been appreciated that clusters are highly efficient at inter-converting light axion-like particles (ALPs) and photons \cite{13053603, 13123947, 14114172} (see \cite{09014085, 09022320, 12070776} for some
earlier work). In this paper, we show that the absence of large distortions of the cluster thermal X-ray bremsstrahlung spectrum may be used to derive the strongest bounds to date on the ALP--photon coupling for light ALPs.

ALPs, reviewed in \cite{12105081}, arise in many theories of physics beyond the Standard Model and are ubiquitous in string theory compactifications.
The phenomenological low-energy Lagrangian for ALPs and photons is given by,
\be
{\cal L} =
\frac{1}{2} \partial_{\mu} a \partial^{\mu} a + \frac{1}{4} F_{\mu \nu} F^{\mu \nu} +
\frac{a}{4 M} F_{\mu \nu} \tilde F^{\mu \nu}
- \frac{1}{2} m_a^2 a^2
\, ,
\label{eq:L}
\ee
where $M$ is the ALP-photon coupling, and $m_a$ denotes the mass of the ALP.

Significant observational and theoretical effort has gone into searching for and constraining such particles.
 The strongest current bound arises
from observations of SN1987A, leading to $M \gtrsim 2 \times 10^{11}\, {\rm GeV}$ \cite{Brockway:1996yr, Grifols:1996id,Payez:2014xsa}.
Planned experiments such as ALPS-II or IAXO are expected to produce competitive bounds \cite{Bahre:2013ywa, Armengaud:2014gea}, e.g.~$M>3\times10^{11}\, {\rm GeV}$ in the IAXO best-case-scenario.

For massive ALPs, a recent paper used the absence of CMB distortions through clusters to produce bounds on $M$ \cite{150702855}, although as
this relies on resonance effects it is only relevant for a small range of ALP masses.

Here, we argue
that current and future data from X-ray observations of galaxy clusters can
significantly improve the bounds on $M$ for the entire small $m_a$ region ($\lesssim 1 \ti 10^{-12}\, {\rm eV}$),
reaching down to around $M \sim 10^{12}$\, GeV.


\subsection{Photon-ALP conversion in galaxy clusters}
As is well known, the third term of the Lagrangian (\ref{eq:L}) induces ALP-photon inter-conversion
in the presence of coherent magnetic fields \cite{Sikivie:1983, Sikivie:1985, Raffelt}. The linearised equations of motion for the modes of energy $\omega$ propagating in the $z$-direction are given by,
\beq
\label{eq:EoM}
\left(\omega + \left(\begin{array}{c c c}
			\Delta_{\gamma} & \Delta_{\rm F} & \Delta_{\gamma a x} \\
			\Delta_{\rm F} & \Delta_{\gamma} & \Delta_{\gamma a y} \\
			\Delta_{\gamma a x} & \Delta_{\gamma a y} & \Delta_{a}
		   \end{array}\right) - i\partial_z\right)\left(\begin{array}{c}
								\Ket{\gamma_x} \\
								\Ket{\gamma_y} \\
								\axion
							      \end{array}\right)= 0 \, ,
\eeq
for the ALP-state $\Ket{a}$ and the two transverse polarisations of the photon beam, $\Ket{\gamma_x}$ and $\Ket{\gamma_y}$.
Here $\Delta_{\gamma} =  -\omega_{\rm pl}^2/2\omega$, $\Delta_{\gamma a i} = B_i/2M$ and $\Delta_a  = -m_a^2/ 2 \omega$.
The plasma frequency of the surrounding medium is given by,
\be
\omega_{\rm pl}  = \sqrt{4\pi\alpha n_e/m_e} = \left( \frac{n_e}{10^{-3} {\rm cm}^{-3}} \right)^{0.5} 1.2 \times 10^{-12}{\rm eV}.
\ee
$\Delta_F$ accounts for Faraday rotation. However this effect depends on wavelength as $\lambda^2$ and thus is negligible at X-ray energies.

For a constant magnetic field in a domain of length $L$, the photon-to-ALP conversion probability for unpolarised light is given by,
\be
\Pga = \half \frac{\z^2}{1+ \z^2} \sin^2 \left( \d \sqrt{1 + \z^2} \right) \, ,
\label{eq:P}
\ee
where $\Theta = \frac{2 B_{\perp} \omega}{M m^2_{eff}}$,
$\Delta = \frac{m_{eff}^2 L}{4 \omega}$,  and $m^2_{eff} = m_a^2 - \omega_{pl}^2$. Here $B_{\perp}$ denotes the component of the magnetic field that is perpendicular to the ALP wave vector. The factor of $\half$ accounts for the fact that only one polarisation state of
light participates in the mixing.

It is instructive to consider the typical values of $\Theta$ and $\Delta$ in galaxy clusters. For $m_a =0$ we have,
\bea
\z &=& 0.28 \left(\frac{B_{\perp}}{1\, {\rm \mu G}} \right)\left(\frac{\omega}{1\, {\rm keV}}\right)\left( \frac{10^{-3}\, {\rm cm^{-3}}}{n_e} \right)\left(\frac{10^{11}\,
{\rm GeV}}{M}\right), \nn \\
\d &=& 0.54 \left( \frac{n_e}{10^{-3}\, {\rm cm^{-3}}} \right) \left( \frac{L}{10\, {\rm kpc}} \right) \left( \frac{1\, {\rm keV}}{\omega} \right).
\label{eq:Delta}
\eea
For $\z, \d \ll 1$, conversion is quadratic in both size and coherence length of the magnetic field,
\be
P_{\gamma \to a} = \frac{1}{2} \z^2 \d^2 = \frac{B^2_{\perp} L^2}{8 M^2} \, ,
\label{eq:P1}
\ee
and ALP-photon conversion is energy-independent. However, for $\z\ll1 $ and $ \d\gg1$, the conversion probability is energy-dependent and
progressively suppressed at lower energies,
\be
P_{\gamma\to a} = \frac{1}{2} \z^2 \sin^2(\d) \sim \frac{m_e^2 B_{\perp}^2}{\pi^2 \alpha^2 n_e^2 M^2} \omega^2 \sin^2\left(\frac{\pi \alpha}{m_e} \frac{n_e L}{\omega} \right) \, .
\label{eq:P2}
\ee
We can define a critical energy marking the crossover between the regimes,
\be
\frac{\omega_{\Delta}}{\rm keV}=0.54\left|\frac{n_e}{10^{-3}\,\text{cm}^{-3}}-\left(\frac{m_a}{1.2\cdot10^{-12}\,\text{eV}}\right)^2\right|\left(\frac{L}{10\,\text{kpc}}\right),
\label{eq:critenergy}
\ee
and thus for typical clusters and for $M \gtrsim 10^{11} {\rm GeV}$, the crossover between regimes occurs within
the range of observed X-ray photon energies. We now show that this can lead to
substantial distortions of the thermal spectrum of the intra-cluster medium, using the Coma cluster as our prime example.


\subsection{X-ray emission from galaxy clusters}

The intracluster medium (ICM) permeating galaxy clusters is a hot thermal plasma
with
temperatures, depending on the cluster, of between 2 and 10 keV.
The cluster is visible in X-rays through the thermal bremsstrahlung of the ICM, with both continuum and line emission.

The continuum emissivity is given by,
\be
I(\nu) = A n_e^2  \frac{ g(\nu, T)  \exp \left(-\frac{h \nu}{kT} \right)}{\sqrt{kT}},
\ee
where $A$ is a constant and the Gaunt factor $g(\nu, T)$ is a slowly varying function of frequency.
The strength of the atomic lines is set by both the overall metallicity and individual abundance of heavy elements, as well as the
local ICM temperature, and can be calculated using programs such as AtomDB \cite{AtomDB}.

Thermal emission from the ICM has been
extensively observed by several generations of X-ray satellites across the last few decades, providing an excellent fit to observations.
 Figure \ref{ComaFull} shows fits to emission from the Coma cluster (adapted from \cite{Arnaud:2000vc} and \cite{nuStar}), which hosts an approximately isothermal ICM with a temperature of $8.1\, {\rm keV}$. Note the excellent quality of the fits, with residuals all below the $10\%$ level.
As a further illustration of the attained precision, we note that the
possible 3.55 keV dark matter line reported in \cite{Bulbul} is
observed as a per cent level effect above the background, thereby
requiring the ability to characterise the background thermal emission to the same level of accuracy.

Thus, galaxy clusters  provide an intrinsic, diffuse, bright and well characterised source of X-ray photons.


\subsection{Cluster spectral distortions from ALPs}

If a significant proportion of the thermal photons were to convert to ALPs, the resulting X-ray spectrum would be distorted.
The photon-to-ALP conversion probability is determined largely by the
structure of the cluster magnetic field.

Generically, clusters support magnetic fields of $\mc{O}(\mu {\rm G})$ strength that are coherent over $1$--$10$ kpc scales (for reviews see \cite{Govoni:2004as, Feretti:1205.1919}). For the Coma cluster, a detailed
multi-scale model of the tangled magnetic field was constructed in \cite{10020594}, and shown to give a good fit to observed Faraday rotation measures.

For the Coma magnetic field of \cite{10020594},
we have simulated photon-to-ALP conversion of the ICM by numerically solving (\ref{eq:EoM})  along a set of sightlines that sample a given field-of-view.
Along any given sightline, photons originate along each point in the cluster, and the net conversion probability is an average  weighted by
 the thermal photon density
 at each point, which scales as $\sim n_e^2$.

The distorted X-ray spectrum is then given by,
\be
f_{\rm distorted}(\omega) = \left(1- \langle P_{\gamma \to a}\rangle (\omega) \right) f_{\rm intrinsic}(\omega) \, ,
\ee
 where $\langle P_{\gamma \to a}\rangle (\omega)$ denotes the
 net conversion probability for the sightlines, averaged
  over the relevant field-of-view.

For $M \lesssim 10^{11}\, {\rm GeV}$,  the conversion probability saturates for the entire X-ray range, 
leading to a uniform reduction of the cluster luminosity to 2/3 of its original amount.
However, as shown in equation \ref{eq:critenergy}, for precisely the most observationally interesting range of $10^{11} < M/{\rm GeV} \lesssim 10^{12}$, the photon-to-ALP conversion probability is large
at higher X-ray energies, while being suppressed at lower energies. Thus, such photon-to-ALP conversion
induces spectral distortions of the thermal ICM spectrum \footnote{We note that this effect is unlikely to generate the cluster soft X-ray excess, as that excess is observed at lower energies.}. In Figure  \ref{fig:resid} we illustrate this by plotting $\langle P_{\gamma\to a}\rangle (\omega)$ obtained from the simulations with $m_a = 0$,  $M=4\times 10^{11}\text{ GeV}$ and a field-of-view of size $(100\,{\rm kpc})^2$ at the centre of the Coma cluster.

\begin{figure*}[ht!]
    \centering
    \begin{subfigure}[t]{0.45\textwidth}
        \centering
        \includegraphics[width=0.85\textwidth]{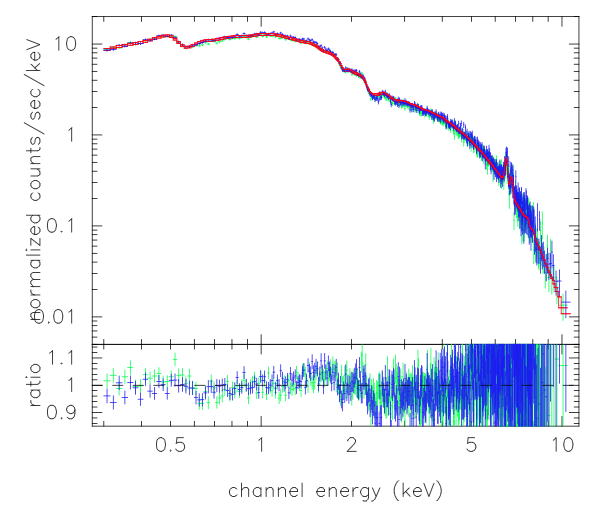}
        \caption{XMM-Newton best fit flux model from Coma, within radius of $\sim 300\kpc$ \cite{Arnaud:2000vc}.}
        \label{fig:spectrum}
    \end{subfigure}%
    ~	~
    \begin{subfigure}[t]{0.45\textwidth}
        \centering
        \includegraphics[width=0.85\textwidth]{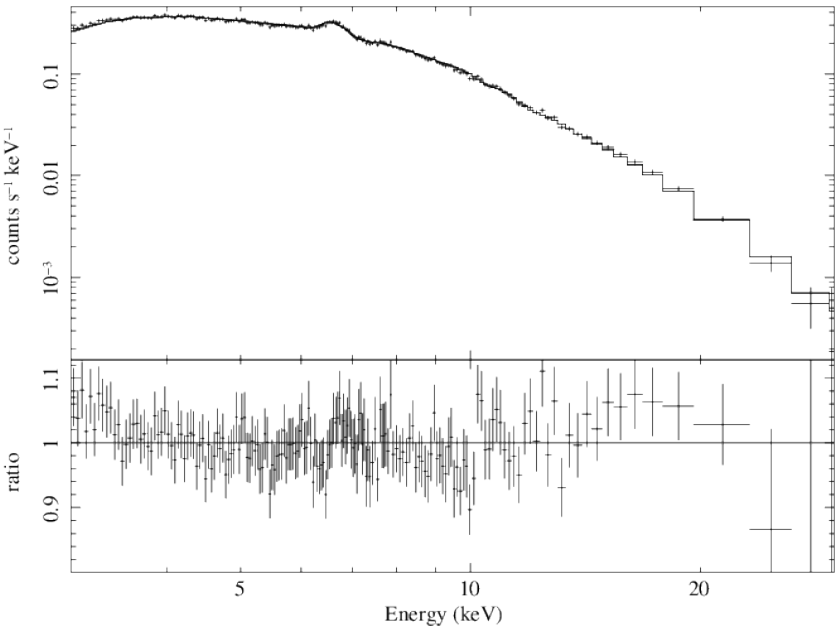}
        \caption{NuSTAR spectrum and best-fit single temperature model for the central $12'\times12'$ of the Coma cluster \cite{nuStar}.}
        \label{fig:resid}
    \end{subfigure}

    \caption{Thermal X-ray spectrum from Coma as seen by XMM-Newton and NuSTAR.}
    \label{ComaFull}
\end{figure*}

\begin{figure*}[t!]
	\centering
	    \centering
    \begin{subfigure}[t]{0.42\textwidth}
        \centering
        \includegraphics[height=2in]{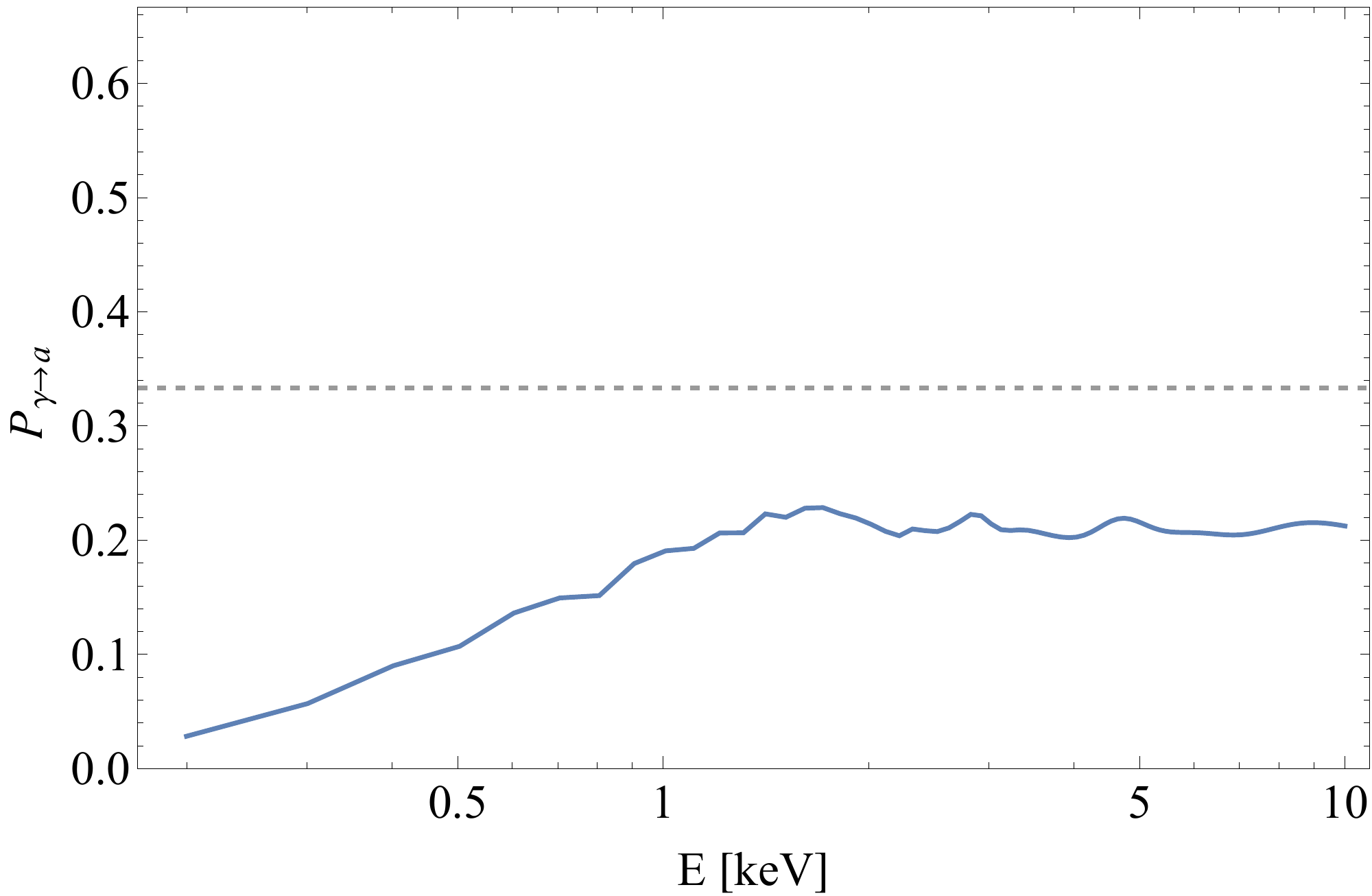}
        \caption{Averaged conversion probabilities for $M=4\times 10^{11}\, {\rm GeV}$, $m_a = 0\, {\rm eV}$, for a 100 kpc field-of-view. The dashed line represents the saturation
        value of 1/3.}
        \label{fig:resid}
    \end{subfigure}
    ~~
      \begin{subfigure}[t]{0.42\textwidth}
        \centering
       \includegraphics[height=2in]{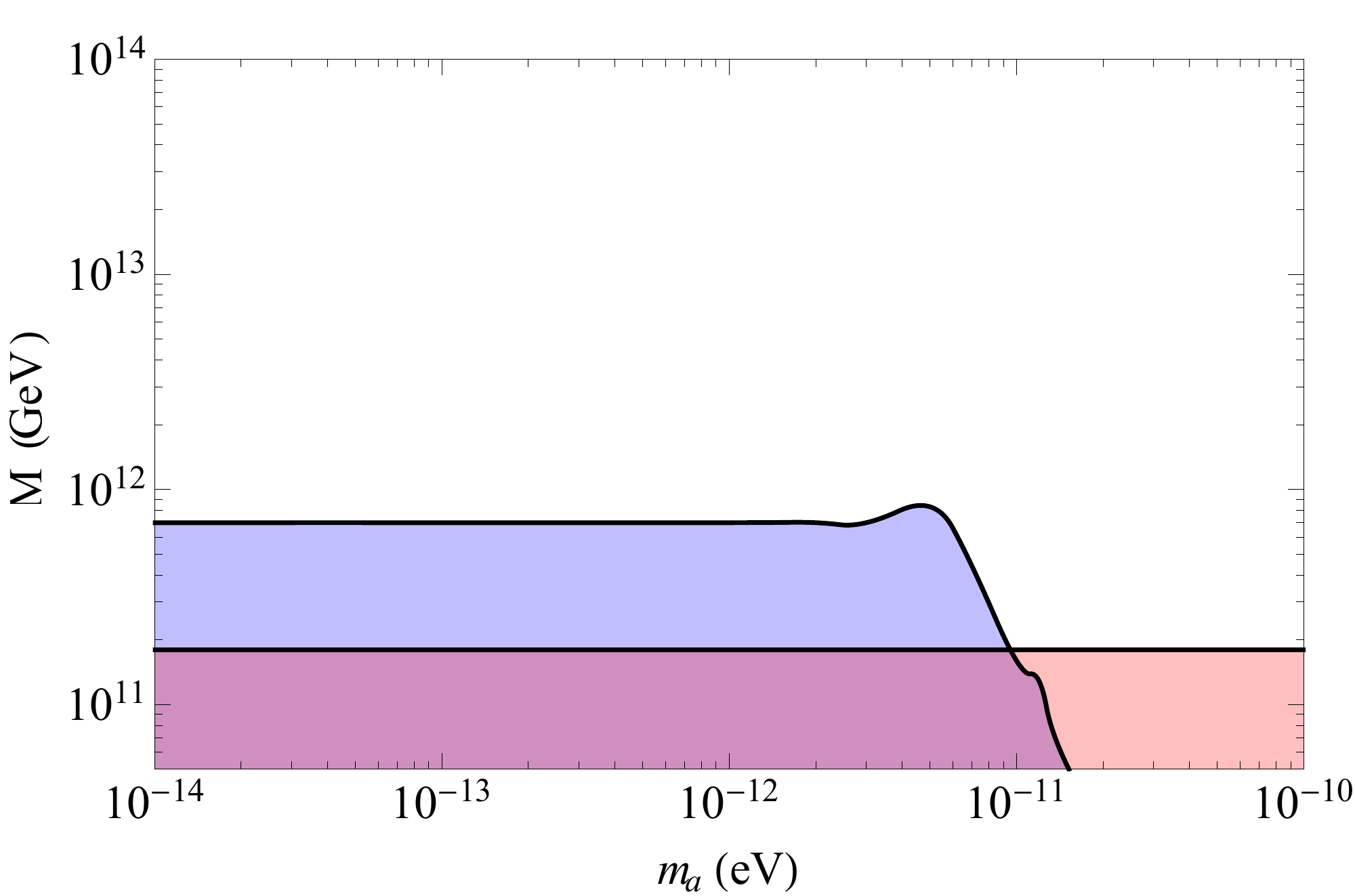}
	\caption{ Astrophysical bounds from the lack of an observation of a gamma-ray burst with SN1987a in red,  bounds on the parameter space from this study in blue. Because of uncertainties in the cluster magnetic field, the precise values of $M$ excluded may be uncertain by a factor of two.}	\label{fig:mM}
    \end{subfigure}
    \caption{Averaged conversion photon-to-ALP conversion probability and
	constraints  on $M$.}
\end{figure*}

Given the quality of the fits in Figure \ref{ComaFull},
the simulated photon-to-ALP conversion probabilities can be used to constrain the ALP parameters
$M$ and $m_a$. By conservatively stipulating the absence of deviations in the spectrum of 10\% or greater,
we find
 the constraints plotted in
   Figure \ref{fig:mM}. For $m_a \lesssim 5\times 10^{-12}\, {\rm eV}$ this provides a bound on the ALP coupling $M\lesssim 7\times 10^{11}\, {\rm GeV}$, while for $m_a \gtrsim 5\times 10^{-12}\, {\rm eV}$, the conversion probability is suppressed by the large ALP mass, and for $m_a \gtrsim 1\times10^{-11}\, {\rm eV}$ the corresponding bound on the coupling becomes weaker than existing astrophysical constraints.

We therefore conclude  that constraints on spectral deviations of the cluster thermal bremsstrahlung spectrum can provide the strongest astrophysical bounds to date on
the ALP-photon coupling, excluding an interesting range of parameters that will be probed in upcoming laboratory experiments.

\subsection{Further properties of the distorted spectrum}

There are several further properties, and thus potentially correlated signals, of an ALP-distorted cluster spectrum.

\begin{figure*}[t!]
	\centering
	 \begin{subfigure}[t]{0.45\textwidth}
		\centering
		\includegraphics[width=0.85\textwidth]{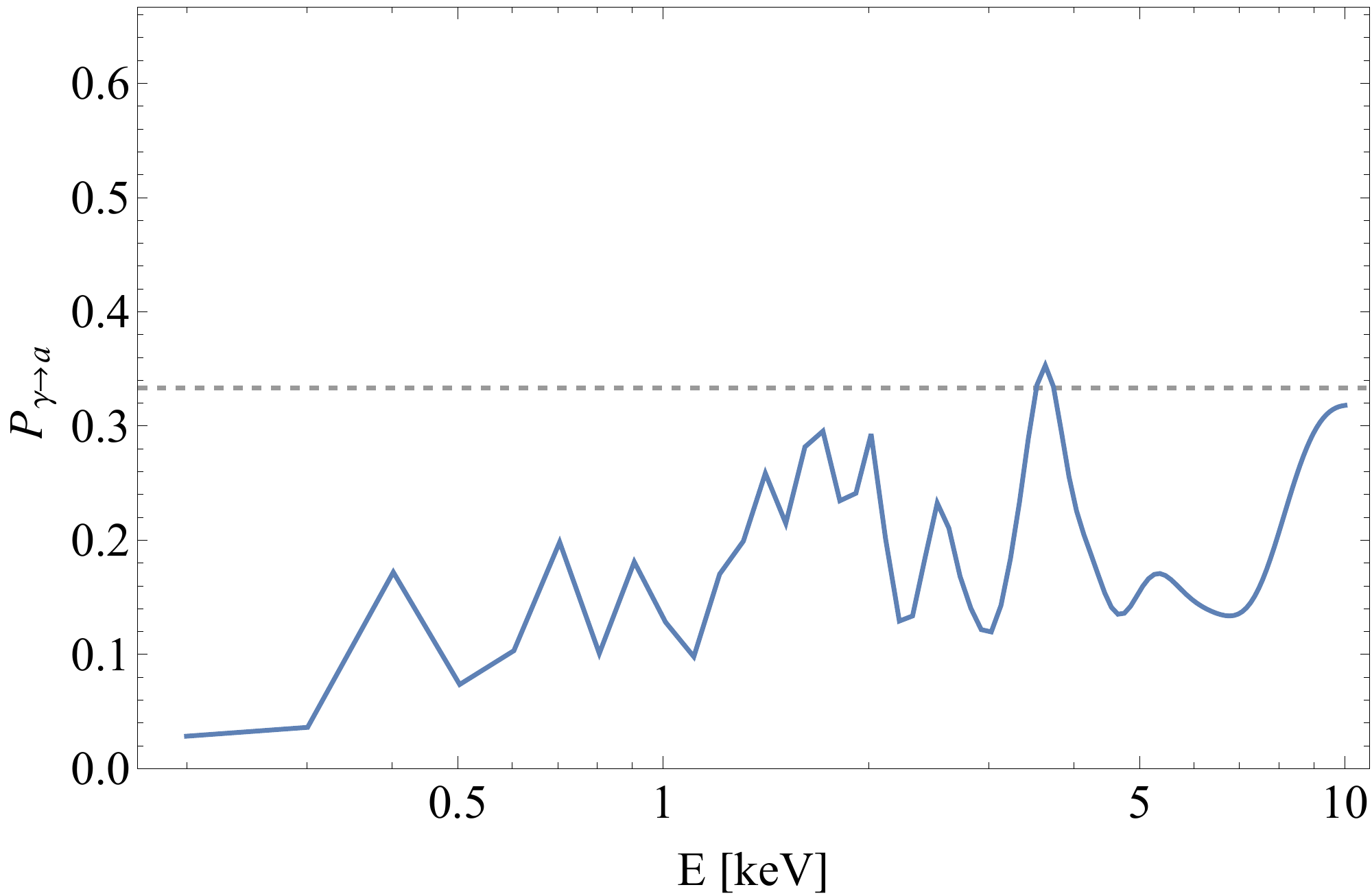}
		\caption{The averaged conversion probabilities.}
		\label{fig:smallfov}
	\end{subfigure}
	~
	 \begin{subfigure}[t]{0.45\textwidth}
		\centering
		\includegraphics[width=0.85\textwidth]{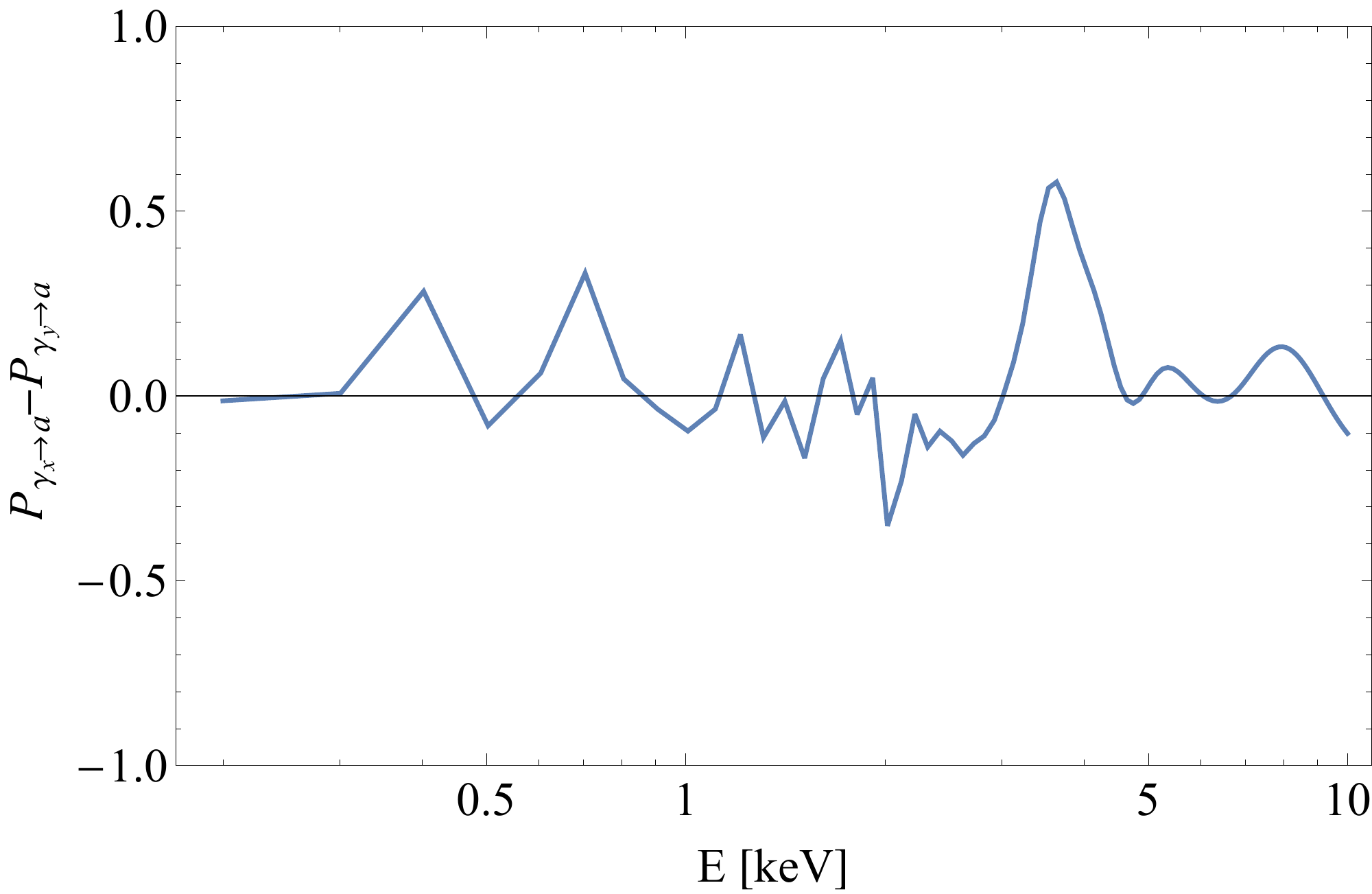}
		\caption{The difference between conversion probabilities for the two polarisations of X-rays, showing the induced polarisation caused by ALP--photon oscillations.}
		\label{fig:pol}
	\end{subfigure}
	\caption{Conversion probabilities and polarisation for a small 5 kpc field-of-view, with coupling $M=4\times 10^{11}\, {\rm GeV}$ and mass $m_a = 0 \, {\rm eV}$.}
\end{figure*}

First, along a single line of sight photon-to-ALP
conversion is highly stochastic and energy-dependent. These features can provide several additional observational signatures, which are observable for small field-of-views.
In Figure \ref{fig:smallfov} we illustrate this by showing the averaged conversion probability for a
$(5\,{\rm kpc})^2$ field-of-view. Since the simulated field is random in nature, the exact energy dependence of the real field will differ, but the presence of these sinusoidal oscillations at X-ray energies is a characteristic feature of ALP-photon conversion, with
the frequency of these oscillations decreasing with increasing energy.

When averaging over a large
field-of-view of size far greater than the coherence length of the cluster magnetic field (as in Figure \ref{fig:resid}), these variations wash out.
However, by extracting the cluster
X-ray spectrum across a small region it may be possible to directly observe such oscillations. The best candidates for this are very nearby clusters, where
the fixed telescope angular resolution corresponds to the smallest physical scales.

For example, the Virgo cluster is at a redshift of $z=0.004$ and there is approximately 500 ks of observation time on Virgo with
the Chandra telescope, whose superb imaging capabilities
gives it arcsecond resolution. By considering spectra extracted from small angular regions instead of the full field-of-view, one
could search for the presence (or absence) of such oscillatory features.

There is another approach to searching for these small scale features.
Clusters are large Mpc-scale objects, and gradients in their internal structure will generally also be large-scale.
In the absence of photon-axion conversion, the bremsstrahlung photon count from nearby pixels on a detector
is expected to be identical, subject to the variations due to Poisson statistics.
Photon-ALP conversion can lead to significant variations in conversion probabilities from nearby sightlines, and so
in the presence of ALPs the photon count from nearby pixels should have a greater variation
than would be expected simply from Poisson statistics.

Furthermore,
along a single line of sight the X-ray emission becomes highly polarised. In Figure \ref{fig:pol} we show that with a small field-of-view this induced polarisation would be observable, but again for larger field-of-views this effect gets washed out.
The result is that for $10^{11}\, {\rm GeV} \lesssim M \lesssim 10^{12}\, {\rm GeV}$, the diffuse thermal ICM emission extracted from a small region would be polarised.
If an X-ray telescope with polarisation capability, such as the European LOFT or Chinese XTP missions, is (finally, after forty years) flown,
the presence or absence of polarised emission from the ICM could lead either to the detection of ALPs or strong bounds on their coupling.
While optical polarisation induced by ALP--photon oscillations has been considered before (see for example  \cite{2012JCAP...07..041P}),
the advantage of a galaxy cluster environment is efficient X-ray conversion and better studied magnetic fields.

The small-scale spectral oscillations can also be sought by considering X-ray point sources that are fortuitously observed through a cluster.
Such a search has been carried out for the Hydra A cluster in \cite{WoutersBrun} using a central AGN source. However this is not an ideal source. Hydra A is at a redshift of $z=0.06$, more than ten times further away than e.g.~the Virgo cluster, and furthermore
the AGN is optically thick, with significant intrinsic absorption at lower X-ray energies, making it harder to separate photon-ALP leakage from this effect.

Another interesting effect would be that the spectral distortions would have a radial dependence within a cluster.
The free electron density decreases with radius, and observations indicate the magnetic field strength also decreases with radial distance from the centre of the cluster.
Photons travelling along off-centred sightlines will then convert at a reduced rate at higher energies (see equation \ref{eq:P1}), and the `step' in the conversion probability appear at lower energies, as implied by equation \ref{eq:critenergy}. This behaviour would be characteristic of ALP-induced distortions.


\subsection{Conclusions and Outlook}

The main point of this paper is that thermal emission from
galaxy clusters provides a well characterised X-ray source, arising within a magnetic field environment that
would lead to significant spectral distortion for ALP masses $m_a \lesssim 1 \ti 10^{-12}\, {\rm eV}$ and
ALP-photon couplings $10^{11}\, {\rm GeV} \lesssim M \lesssim 7 \times 10^{11}\, {\rm GeV}$.
As such spectral distortions do not appear to be observed, this places highly competitive, and potentially the most stringent, bounds
on ALP parameter space. We should of course also caution that bounds on $M$ can only ever be as good as
knowledge of the astrophysical magnetic field, which for clusters is probably currently uncertain up to a factor of two.

We have used the absence of any reported large deviations from a thermal spectrum to bound the ALP parameter space. Existing analyses however are not optimised for ALP searches. For these, it is beneficial to extract spectra over a reduced field-of-view and search for
 energy-dependent oscillations.
We plan to pursue this approach in future.

An attractive reason to study X-ray emission from clusters and its connection to ALPs is that the data is only going to get better.
The three current satellites XMM-Newton, Chandra, and Suzaku, all use CCD detectors with
an intrinsic resolution of $\Delta E \sim 100\, {\rm eV}$.
Next year ASTRO-H will launch, which will allow a far greater precision to be obtained due to
the 7 eV resolution available from the microcalorimeters on board its Soft X-ray Spectrometer.
This improved resolution is highly relevant for searches for the sinusoidal spectral
oscillations characteristic of ALPs. Over the longer term, ATHENA has been approved by ESA
for a 2028 launch.

Knowledge of cluster magnetic fields will also be enhanced over the short, medium and longer term. Through observing
(or not) upscattered Inverse Compton-CMB photons, hard X-ray imaging telescopes will either produce direct measurements of the magnetic field,
or more stringent lower bounds on the field strength. Over the next decade, the Square Kilometer Array will also come online.
With a large increase in the number of available radio sources, this will improve Faraday rotation measurements of galaxy clusters' magnetic fields.

\subsection*{Acknowledgements}

We thank Pedro Alzarez, Francesca Day, Nick Jennings, Sven Krippendorf, Markus Rummel for conversations.
This project is funded in part by the European Research Council as Starting Grant 307605-SUSYBREAKING.
JC is also funded by a Royal Society University Research Fellowship.

\bibliographystyle{apsrev4-1}
\bibliography{Xrayspec}

\begin{thebibliography}{26}%
\makeatletter
\providecommand \@ifxundefined [1]{%
 \@ifx{#1\undefined}
}%
\providecommand \@ifnum [1]{%
 \ifnum #1\expandafter \@firstoftwo
 \else \expandafter \@secondoftwo
 \fi
}%
\providecommand \@ifx [1]{%
 \ifx #1\expandafter \@firstoftwo
 \else \expandafter \@secondoftwo
 \fi
}%
\providecommand \natexlab [1]{#1}%
\providecommand \enquote  [1]{``#1''}%
\providecommand \bibnamefont  [1]{#1}%
\providecommand \bibfnamefont [1]{#1}%
\providecommand \citenamefont [1]{#1}%
\providecommand \href@noop [0]{\@secondoftwo}%
\providecommand \href [0]{\begingroup \@sanitize@url \@href}%
\providecommand \@href[1]{\@@startlink{#1}\@@href}%
\providecommand \@@href[1]{\endgroup#1\@@endlink}%
\providecommand \@sanitize@url [0]{\catcode `\\12\catcode `\$12\catcode
  `\&12\catcode `\#12\catcode `\^12\catcode `\_12\catcode `\%12\relax}%
\providecommand \@@startlink[1]{}%
\providecommand \@@endlink[0]{}%
\providecommand \url  [0]{\begingroup\@sanitize@url \@url }%
\providecommand \@url [1]{\endgroup\@href {#1}{\urlprefix }}%
\providecommand \urlprefix  [0]{URL }%
\providecommand \Eprint [0]{\href }%
\providecommand \doibase [0]{http://dx.doi.org/}%
\providecommand \selectlanguage [0]{\@gobble}%
\providecommand \bibinfo  [0]{\@secondoftwo}%
\providecommand \bibfield  [0]{\@secondoftwo}%
\providecommand \translation [1]{[#1]}%
\providecommand \BibitemOpen [0]{}%
\providecommand \bibitemStop [0]{}%
\providecommand \bibitemNoStop [0]{.\EOS\space}%
\providecommand \EOS [0]{\spacefactor3000\relax}%
\providecommand \BibitemShut  [1]{\csname bibitem#1\endcsname}%
\let\auto@bib@innerbib\@empty
\bibitem [{\citenamefont {Conlon}\ and\ \citenamefont
  {Marsh}(2013)}]{13053603}%
  \BibitemOpen
  \bibfield  {author} {\bibinfo {author} {\bibfnamefont {J.~P.}\ \bibnamefont
  {Conlon}}\ and\ \bibinfo {author} {\bibfnamefont {M.~C.~D.}\ \bibnamefont
  {Marsh}},\ }\href {\doibase 10.1103/PhysRevLett.111.151301} {\bibfield
  {journal} {\bibinfo  {journal} {Phys. Rev. Lett.}\ }\textbf {\bibinfo
  {volume} {111}},\ \bibinfo {pages} {151301} (\bibinfo {year} {2013})},\
  \Eprint {http://arxiv.org/abs/1305.3603} {arXiv:1305.3603 [astro-ph.CO]}
  \BibitemShut {NoStop}%
\bibitem [{\citenamefont {Angus}\ \emph {et~al.}(2014)\citenamefont {Angus},
  \citenamefont {Conlon}, \citenamefont {Marsh}, \citenamefont {Powell},\ and\
  \citenamefont {Witkowski}}]{13123947}%
  \BibitemOpen
  \bibfield  {author} {\bibinfo {author} {\bibfnamefont {S.}~\bibnamefont
  {Angus}}, \bibinfo {author} {\bibfnamefont {J.~P.}\ \bibnamefont {Conlon}},
  \bibinfo {author} {\bibfnamefont {M.~C.~D.}\ \bibnamefont {Marsh}}, \bibinfo
  {author} {\bibfnamefont {A.~J.}\ \bibnamefont {Powell}}, \ and\ \bibinfo
  {author} {\bibfnamefont {L.~T.}\ \bibnamefont {Witkowski}},\ }\href {\doibase
  10.1088/1475-7516/2014/09/026} {\bibfield  {journal} {\bibinfo  {journal}
  {JCAP}\ }\textbf {\bibinfo {volume} {1409}},\ \bibinfo {pages} {026}
  (\bibinfo {year} {2014})},\ \Eprint {http://arxiv.org/abs/1312.3947}
  {arXiv:1312.3947 [astro-ph.HE]} \BibitemShut {NoStop}%
\bibitem [{\citenamefont {Powell}(2015)}]{14114172}%
  \BibitemOpen
  \bibfield  {author} {\bibinfo {author} {\bibfnamefont {A.~J.}\ \bibnamefont
  {Powell}},\ }\href {\doibase 10.1088/1475-7516/2015/09/017} {\bibfield
  {journal} {\bibinfo  {journal} {JCAP}\ }\textbf {\bibinfo {volume} {1509}},\
  \bibinfo {pages} {017} (\bibinfo {year} {2015})},\ \Eprint
  {http://arxiv.org/abs/1411.4172} {arXiv:1411.4172 [astro-ph.CO]} \BibitemShut
  {NoStop}%
\bibitem [{\citenamefont {Fairbairn}\ \emph {et~al.}(2011)\citenamefont
  {Fairbairn}, \citenamefont {Rashba},\ and\ \citenamefont
  {Troitsky}}]{09014085}%
  \BibitemOpen
  \bibfield  {author} {\bibinfo {author} {\bibfnamefont {M.}~\bibnamefont
  {Fairbairn}}, \bibinfo {author} {\bibfnamefont {T.}~\bibnamefont {Rashba}}, \
  and\ \bibinfo {author} {\bibfnamefont {S.~V.}\ \bibnamefont {Troitsky}},\
  }\href {\doibase 10.1103/PhysRevD.84.125019} {\bibfield  {journal} {\bibinfo
  {journal} {Phys. Rev.}\ }\textbf {\bibinfo {volume} {D84}},\ \bibinfo {pages}
  {125019} (\bibinfo {year} {2011})},\ \Eprint {http://arxiv.org/abs/0901.4085}
  {arXiv:0901.4085 [astro-ph.HE]} \BibitemShut {NoStop}%
\bibitem [{\citenamefont {Burrage}\ \emph {et~al.}(2009)\citenamefont
  {Burrage}, \citenamefont {Davis},\ and\ \citenamefont {Shaw}}]{09022320}%
  \BibitemOpen
  \bibfield  {author} {\bibinfo {author} {\bibfnamefont {C.}~\bibnamefont
  {Burrage}}, \bibinfo {author} {\bibfnamefont {A.-C.}\ \bibnamefont {Davis}},
  \ and\ \bibinfo {author} {\bibfnamefont {D.~J.}\ \bibnamefont {Shaw}},\
  }\href {\doibase 10.1103/PhysRevLett.102.201101} {\bibfield  {journal}
  {\bibinfo  {journal} {Phys. Rev. Lett.}\ }\textbf {\bibinfo {volume} {102}},\
  \bibinfo {pages} {201101} (\bibinfo {year} {2009})},\ \Eprint
  {http://arxiv.org/abs/0902.2320} {arXiv:0902.2320 [astro-ph.CO]} \BibitemShut
  {NoStop}%
\bibitem [{\citenamefont {Horns}\ \emph {et~al.}(2012)\citenamefont {Horns},
  \citenamefont {Maccione}, \citenamefont {Meyer}, \citenamefont {Mirizzi},
  \citenamefont {Montanino},\ and\ \citenamefont {Roncadelli}}]{12070776}%
  \BibitemOpen
  \bibfield  {author} {\bibinfo {author} {\bibfnamefont {D.}~\bibnamefont
  {Horns}}, \bibinfo {author} {\bibfnamefont {L.}~\bibnamefont {Maccione}},
  \bibinfo {author} {\bibfnamefont {M.}~\bibnamefont {Meyer}}, \bibinfo
  {author} {\bibfnamefont {A.}~\bibnamefont {Mirizzi}}, \bibinfo {author}
  {\bibfnamefont {D.}~\bibnamefont {Montanino}}, \ and\ \bibinfo {author}
  {\bibfnamefont {M.}~\bibnamefont {Roncadelli}},\ }\href {\doibase
  10.1103/PhysRevD.86.075024} {\bibfield  {journal} {\bibinfo  {journal} {Phys.
  Rev.}\ }\textbf {\bibinfo {volume} {D86}},\ \bibinfo {pages} {075024}
  (\bibinfo {year} {2012})},\ \Eprint {http://arxiv.org/abs/1207.0776}
  {arXiv:1207.0776 [astro-ph.HE]} \BibitemShut {NoStop}%
\bibitem [{\citenamefont {Ringwald}(2012)}]{12105081}%
  \BibitemOpen
  \bibfield  {author} {\bibinfo {author} {\bibfnamefont {A.}~\bibnamefont
  {Ringwald}},\ }\href {\doibase 10.1016/j.dark.2012.10.008} {\bibfield
  {journal} {\bibinfo  {journal} {Phys.Dark Univ.}\ }\textbf {\bibinfo {volume}
  {1}},\ \bibinfo {pages} {116} (\bibinfo {year} {2012})},\ \Eprint
  {http://arxiv.org/abs/1210.5081} {arXiv:1210.5081 [hep-ph]} \BibitemShut
  {NoStop}%
\bibitem [{\citenamefont {Brockway}\ \emph {et~al.}(1996)\citenamefont
  {Brockway}, \citenamefont {Carlson},\ and\ \citenamefont
  {Raffelt}}]{Brockway:1996yr}%
  \BibitemOpen
  \bibfield  {author} {\bibinfo {author} {\bibfnamefont {J.~W.}\ \bibnamefont
  {Brockway}}, \bibinfo {author} {\bibfnamefont {E.~D.}\ \bibnamefont
  {Carlson}}, \ and\ \bibinfo {author} {\bibfnamefont {G.~G.}\ \bibnamefont
  {Raffelt}},\ }\href {\doibase 10.1016/0370-2693(96)00778-2} {\bibfield
  {journal} {\bibinfo  {journal} {Phys.Lett.}\ }\textbf {\bibinfo {volume}
  {B383}},\ \bibinfo {pages} {439} (\bibinfo {year} {1996})},\ \Eprint
  {http://arxiv.org/abs/astro-ph/9605197} {arXiv:astro-ph/9605197 [astro-ph]}
  \BibitemShut {NoStop}%
\bibitem [{\citenamefont {Grifols}\ \emph {et~al.}(1996)\citenamefont
  {Grifols}, \citenamefont {Masso},\ and\ \citenamefont
  {Toldra}}]{Grifols:1996id}%
  \BibitemOpen
  \bibfield  {author} {\bibinfo {author} {\bibfnamefont {J.}~\bibnamefont
  {Grifols}}, \bibinfo {author} {\bibfnamefont {E.}~\bibnamefont {Masso}}, \
  and\ \bibinfo {author} {\bibfnamefont {R.}~\bibnamefont {Toldra}},\ }\href
  {\doibase 10.1103/PhysRevLett.77.2372} {\bibfield  {journal} {\bibinfo
  {journal} {Phys.Rev.Lett.}\ }\textbf {\bibinfo {volume} {77}},\ \bibinfo
  {pages} {2372} (\bibinfo {year} {1996})},\ \Eprint
  {http://arxiv.org/abs/astro-ph/9606028} {arXiv:astro-ph/9606028 [astro-ph]}
  \BibitemShut {NoStop}%
\bibitem [{\citenamefont {Payez}\ \emph {et~al.}(2014)\citenamefont {Payez},
  \citenamefont {Evoli}, \citenamefont {Fischer}, \citenamefont {Giannotti},
  \citenamefont {Mirizzi} \emph {et~al.}}]{Payez:2014xsa}%
  \BibitemOpen
  \bibfield  {author} {\bibinfo {author} {\bibfnamefont {A.}~\bibnamefont
  {Payez}}, \bibinfo {author} {\bibfnamefont {C.}~\bibnamefont {Evoli}},
  \bibinfo {author} {\bibfnamefont {T.}~\bibnamefont {Fischer}}, \bibinfo
  {author} {\bibfnamefont {M.}~\bibnamefont {Giannotti}}, \bibinfo {author}
  {\bibfnamefont {A.}~\bibnamefont {Mirizzi}},  \emph {et~al.},\ }\href@noop {}
  {\  (\bibinfo {year} {2014})},\ \Eprint {http://arxiv.org/abs/1410.3747}
  {arXiv:1410.3747 [astro-ph.HE]} \BibitemShut {NoStop}%
\bibitem [{\citenamefont {B{\"a}hre}\ \emph {et~al.}(2013)\citenamefont
  {B{\"a}hre}, \citenamefont {D{\"o}brich}, \citenamefont
  {Dreyling-Eschweiler}, \citenamefont {Ghazaryan}, \citenamefont {Hodajerdi}
  \emph {et~al.}}]{Bahre:2013ywa}%
  \BibitemOpen
  \bibfield  {author} {\bibinfo {author} {\bibfnamefont {R.}~\bibnamefont
  {B{\"a}hre}}, \bibinfo {author} {\bibfnamefont {B.}~\bibnamefont
  {D{\"o}brich}}, \bibinfo {author} {\bibfnamefont {J.}~\bibnamefont
  {Dreyling-Eschweiler}}, \bibinfo {author} {\bibfnamefont {S.}~\bibnamefont
  {Ghazaryan}}, \bibinfo {author} {\bibfnamefont {R.}~\bibnamefont
  {Hodajerdi}},  \emph {et~al.},\ }\href {\doibase
  10.1088/1748-0221/8/09/T09001} {\bibfield  {journal} {\bibinfo  {journal}
  {JINST}\ }\textbf {\bibinfo {volume} {8}},\ \bibinfo {pages} {T09001}
  (\bibinfo {year} {2013})},\ \Eprint {http://arxiv.org/abs/1302.5647}
  {arXiv:1302.5647 [physics.ins-det]} \BibitemShut {NoStop}%
\bibitem [{\citenamefont {Armengaud}\ \emph {et~al.}(2014)\citenamefont
  {Armengaud}, \citenamefont {Avignone}, \citenamefont {Betz}, \citenamefont
  {Brax}, \citenamefont {Brun} \emph {et~al.}}]{Armengaud:2014gea}%
  \BibitemOpen
  \bibfield  {author} {\bibinfo {author} {\bibfnamefont {E.}~\bibnamefont
  {Armengaud}}, \bibinfo {author} {\bibfnamefont {F.}~\bibnamefont {Avignone}},
  \bibinfo {author} {\bibfnamefont {M.}~\bibnamefont {Betz}}, \bibinfo {author}
  {\bibfnamefont {P.}~\bibnamefont {Brax}}, \bibinfo {author} {\bibfnamefont
  {P.}~\bibnamefont {Brun}},  \emph {et~al.},\ }\href {\doibase
  10.1088/1748-0221/9/05/T05002} {\bibfield  {journal} {\bibinfo  {journal}
  {JINST}\ }\textbf {\bibinfo {volume} {9}},\ \bibinfo {pages} {T05002}
  (\bibinfo {year} {2014})},\ \Eprint {http://arxiv.org/abs/1401.3233}
  {arXiv:1401.3233 [physics.ins-det]} \BibitemShut {NoStop}%
\bibitem [{\citenamefont {Schlederer}\ and\ \citenamefont
  {Sigl}(2015)}]{150702855}%
  \BibitemOpen
  \bibfield  {author} {\bibinfo {author} {\bibfnamefont {M.}~\bibnamefont
  {Schlederer}}\ and\ \bibinfo {author} {\bibfnamefont {G.}~\bibnamefont
  {Sigl}},\ }\href@noop {} {\  (\bibinfo {year} {2015})},\ \Eprint
  {http://arxiv.org/abs/1507.02855} {arXiv:1507.02855 [hep-ph]} \BibitemShut
  {NoStop}%
\bibitem [{\citenamefont {Sikivie}(1983)}]{Sikivie:1983}%
  \BibitemOpen
  \bibfield  {author} {\bibinfo {author} {\bibfnamefont {P.}~\bibnamefont
  {Sikivie}},\ }\href {\doibase 10.1103/PhysRevLett.51.1415} {\bibfield
  {journal} {\bibinfo  {journal} {Phys.Rev.Lett.}\ }\textbf {\bibinfo {volume}
  {51}},\ \bibinfo {pages} {1415} (\bibinfo {year} {1983})}\BibitemShut
  {NoStop}%
\bibitem [{\citenamefont {Sikivie}(1985)}]{Sikivie:1985}%
  \BibitemOpen
  \bibfield  {author} {\bibinfo {author} {\bibfnamefont {P.}~\bibnamefont
  {Sikivie}},\ }\href {\doibase 10.1103/PhysRevD.36.974,
  10.1103/PhysRevD.32.2988} {\bibfield  {journal} {\bibinfo  {journal}
  {Phys.Rev.}\ }\textbf {\bibinfo {volume} {D32}},\ \bibinfo {pages} {2988}
  (\bibinfo {year} {1985})}\BibitemShut {NoStop}%
\bibitem [{\citenamefont {{Raffelt}}\ and\ \citenamefont
  {{Stodolsky}}(1988)}]{Raffelt}%
  \BibitemOpen
  \bibfield  {author} {\bibinfo {author} {\bibfnamefont {G.}~\bibnamefont
  {{Raffelt}}}\ and\ \bibinfo {author} {\bibfnamefont {L.}~\bibnamefont
  {{Stodolsky}}},\ }\href {\doibase 10.1103/PhysRevD.37.1237} {\bibfield
  {journal} {\bibinfo  {journal} {Phys. Rev.}\ }\textbf {\bibinfo {volume}
  {D37}},\ \bibinfo {pages} {1237} (\bibinfo {year} {1988})}\BibitemShut
  {NoStop}%
\bibitem [{Ato()}]{AtomDB}%
  \BibitemOpen
  \href@noop {} {\enquote {\bibinfo {title} {Atomdb},}\ }\bibinfo
  {howpublished} {\url{http://www.atomdb.org}}\BibitemShut {NoStop}%
\bibitem [{\citenamefont {Arnaud}\ \emph {et~al.}(2001)\citenamefont {Arnaud},
  \citenamefont {Aghanim}, \citenamefont {Gastaud}, \citenamefont {Neumann},
  \citenamefont {Lumb} \emph {et~al.}}]{Arnaud:2000vc}%
  \BibitemOpen
  \bibfield  {author} {\bibinfo {author} {\bibfnamefont {M.}~\bibnamefont
  {Arnaud}}, \bibinfo {author} {\bibfnamefont {N.}~\bibnamefont {Aghanim}},
  \bibinfo {author} {\bibfnamefont {R.}~\bibnamefont {Gastaud}}, \bibinfo
  {author} {\bibfnamefont {D.}~\bibnamefont {Neumann}}, \bibinfo {author}
  {\bibfnamefont {D.}~\bibnamefont {Lumb}},  \emph {et~al.},\ }\href {\doibase
  10.1051/0004-6361:20000195} {\bibfield  {journal} {\bibinfo  {journal}
  {Astron.Astrophys.}\ }\textbf {\bibinfo {volume} {365}},\ \bibinfo {pages}
  {L67} (\bibinfo {year} {2001})},\ \Eprint
  {http://arxiv.org/abs/astro-ph/0011086} {arXiv:astro-ph/0011086 [astro-ph]}
  \BibitemShut {NoStop}%
\bibitem [{\citenamefont {Gastaldello}\ \emph {et~al.}(2015)\citenamefont
  {Gastaldello} \emph {et~al.}}]{nuStar}%
  \BibitemOpen
  \bibfield  {author} {\bibinfo {author} {\bibfnamefont {F.}~\bibnamefont
  {Gastaldello}} \emph {et~al.},\ }\href {\doibase 10.1088/0004-637X/800/2/139}
  {\bibfield  {journal} {\bibinfo  {journal} {Astrophys. J.}\ }\textbf
  {\bibinfo {volume} {800}},\ \bibinfo {pages} {139} (\bibinfo {year}
  {2015})},\ \Eprint {http://arxiv.org/abs/1411.1573} {arXiv:1411.1573
  [astro-ph.HE]} \BibitemShut {NoStop}%
\bibitem [{\citenamefont {Bulbul}\ \emph {et~al.}(2014)\citenamefont {Bulbul},
  \citenamefont {Markevitch}, \citenamefont {Foster}, \citenamefont {Smith},
  \citenamefont {Loewenstein},\ and\ \citenamefont {Randall}}]{Bulbul}%
  \BibitemOpen
  \bibfield  {author} {\bibinfo {author} {\bibfnamefont {E.}~\bibnamefont
  {Bulbul}}, \bibinfo {author} {\bibfnamefont {M.}~\bibnamefont {Markevitch}},
  \bibinfo {author} {\bibfnamefont {A.}~\bibnamefont {Foster}}, \bibinfo
  {author} {\bibfnamefont {R.~K.}\ \bibnamefont {Smith}}, \bibinfo {author}
  {\bibfnamefont {M.}~\bibnamefont {Loewenstein}}, \ and\ \bibinfo {author}
  {\bibfnamefont {S.~W.}\ \bibnamefont {Randall}},\ }\href {\doibase
  10.1088/0004-637X/789/1/13} {\bibfield  {journal} {\bibinfo  {journal}
  {Astrophys. J.}\ }\textbf {\bibinfo {volume} {789}},\ \bibinfo {pages} {13}
  (\bibinfo {year} {2014})},\ \Eprint {http://arxiv.org/abs/1402.2301}
  {arXiv:1402.2301 [astro-ph.CO]} \BibitemShut {NoStop}%
\bibitem [{\citenamefont {Govoni}\ and\ \citenamefont
  {Feretti}(2004)}]{Govoni:2004as}%
  \BibitemOpen
  \bibfield  {author} {\bibinfo {author} {\bibfnamefont {F.}~\bibnamefont
  {Govoni}}\ and\ \bibinfo {author} {\bibfnamefont {L.}~\bibnamefont
  {Feretti}},\ }\href {\doibase 10.1142/S0218271804005080} {\bibfield
  {journal} {\bibinfo  {journal} {Int.J.Mod.Phys.}\ }\textbf {\bibinfo {volume}
  {D13}},\ \bibinfo {pages} {1549} (\bibinfo {year} {2004})},\ \Eprint
  {http://arxiv.org/abs/astro-ph/0410182} {arXiv:astro-ph/0410182 [astro-ph]}
  \BibitemShut {NoStop}%
\bibitem [{\citenamefont {{Feretti}}\ \emph {et~al.}(2012)\citenamefont
  {{Feretti}}, \citenamefont {{Giovannini}}, \citenamefont {{Govoni}},\ and\
  \citenamefont {{Murgia}}}]{Feretti:1205.1919}%
  \BibitemOpen
  \bibfield  {author} {\bibinfo {author} {\bibfnamefont {L.}~\bibnamefont
  {{Feretti}}}, \bibinfo {author} {\bibfnamefont {G.}~\bibnamefont
  {{Giovannini}}}, \bibinfo {author} {\bibfnamefont {F.}~\bibnamefont
  {{Govoni}}}, \ and\ \bibinfo {author} {\bibfnamefont {M.}~\bibnamefont
  {{Murgia}}},\ }\href {\doibase 10.1007/s00159-012-0054-z} {\bibfield
  {journal} {\bibinfo  {journal} {Astron. Astrophys. Rev.}\ }\textbf {\bibinfo
  {volume} {20}},\ \bibinfo {pages} {54} (\bibinfo {year} {2012})},\ \Eprint
  {http://arxiv.org/abs/1205.1919} {arXiv:1205.1919 [astro-ph.CO]} \BibitemShut
  {NoStop}%
\bibitem [{\citenamefont {{Bonafede}}\ \emph {et~al.}(2010)\citenamefont
  {{Bonafede}}, \citenamefont {{Feretti}}, \citenamefont {{Murgia}},
  \citenamefont {{Govoni}}, \citenamefont {{Giovannini}}, \citenamefont
  {{Dallacasa}}, \citenamefont {{Dolag}},\ and\ \citenamefont
  {{Taylor}}}]{10020594}%
  \BibitemOpen
  \bibfield  {author} {\bibinfo {author} {\bibfnamefont {A.}~\bibnamefont
  {{Bonafede}}}, \bibinfo {author} {\bibfnamefont {L.}~\bibnamefont
  {{Feretti}}}, \bibinfo {author} {\bibfnamefont {M.}~\bibnamefont {{Murgia}}},
  \bibinfo {author} {\bibfnamefont {F.}~\bibnamefont {{Govoni}}}, \bibinfo
  {author} {\bibfnamefont {G.}~\bibnamefont {{Giovannini}}}, \bibinfo {author}
  {\bibfnamefont {D.}~\bibnamefont {{Dallacasa}}}, \bibinfo {author}
  {\bibfnamefont {K.}~\bibnamefont {{Dolag}}}, \ and\ \bibinfo {author}
  {\bibfnamefont {G.~B.}\ \bibnamefont {{Taylor}}},\ }\href {\doibase
  10.1051/0004-6361/200913696} {\bibfield  {journal} {\bibinfo  {journal}
  {Astron. Astrophys.}\ }\textbf {\bibinfo {volume} {513}},\ \bibinfo {eid}
  {A30} (\bibinfo {year} {2010})},\ \Eprint {http://arxiv.org/abs/1002.0594}
  {arXiv:1002.0594 [astro-ph.CO]} \BibitemShut {NoStop}%
\bibitem [{Note1()}]{Note1}%
  \BibitemOpen
  \bibinfo {note} {We note it seems unlikely this distortion could generate the
  cluster soft excess, as that is observed at lower energies,}\BibitemShut
  {NoStop}%
\bibitem [{\citenamefont {{Payez}}\ \emph {et~al.}(2012)\citenamefont
  {{Payez}}, \citenamefont {{Cudell}},\ and\ \citenamefont
  {{Hutsem{\'e}kers}}}]{2012JCAP...07..041P}%
  \BibitemOpen
  \bibfield  {author} {\bibinfo {author} {\bibfnamefont {A.}~\bibnamefont
  {{Payez}}}, \bibinfo {author} {\bibfnamefont {J.~R.}\ \bibnamefont
  {{Cudell}}}, \ and\ \bibinfo {author} {\bibfnamefont {D.}~\bibnamefont
  {{Hutsem{\'e}kers}}},\ }\href {\doibase 10.1088/1475-7516/2012/07/041}
  {\bibfield  {journal} {\bibinfo  {journal} {JCAP}\ }\textbf {\bibinfo
  {volume} {7}},\ \bibinfo {eid} {041} (\bibinfo {year} {2012})},\ \Eprint
  {http://arxiv.org/abs/1204.6187} {arXiv:1204.6187 [astro-ph.CO]} \BibitemShut
  {NoStop}%
\bibitem [{\citenamefont {Wouters}\ and\ \citenamefont
  {Brun}(2013)}]{WoutersBrun}%
  \BibitemOpen
  \bibfield  {author} {\bibinfo {author} {\bibfnamefont {D.}~\bibnamefont
  {Wouters}}\ and\ \bibinfo {author} {\bibfnamefont {P.}~\bibnamefont {Brun}},\
  }\href {\doibase 10.1088/0004-637X/772/1/44} {\bibfield  {journal} {\bibinfo
  {journal} {Astrophys. J.}\ }\textbf {\bibinfo {volume} {772}},\ \bibinfo
  {pages} {44} (\bibinfo {year} {2013})},\ \Eprint
  {http://arxiv.org/abs/1304.0989} {arXiv:1304.0989 [astro-ph.HE]} \BibitemShut
  {NoStop}%
\end{thebibliography}%

\end{document}